\documentclass[reprint,prx,superscriptaddress,nofootinbib]{revtex4-1}

\usepackage{multirow}
\usepackage{epsfig}
\usepackage{color}
\usepackage{amsmath}
\usepackage{amssymb}
\usepackage{longtable}
\usepackage{dcolumn}\usepackage{bm}\usepackage{ulem}
\usepackage{float}
\usepackage{multirow}\usepackage{hhline}

\providecommand{\abs}[1]{\lvert#1\rvert}

\newcommand{\hJpart}[2]{#1^{(#2)}}
\newcommand{\hp}{\hJpart{h}{p}}
\newcommand{\hb}{\hJpart{h}{b}}
\newcommand{\Jp}{\hJpart{J}{p}}
\newcommand{\Jb}{\hJpart{J}{b}}
\newcommand{\hbi}[1]{\hJpart{h}{b,#1}}
\newcommand{\hpi}[1]{\hJpart{h}{p,#1}}
\newcommand{\Jbi}[1]{\hJpart{J}{b,#1}}
\newcommand{\Jpi}[1]{\hJpart{J}{p,#1}}

\draft 
\begin{document}

\title{Determination and correction of persistent biases in quantum annealers}

\author{Alejandro Perdomo-Ortiz\footnotemark}
\email[Corresponding author's e-mail: ]{alejandro.perdomoortiz@nasa.gov}
\affiliation{Quantum Artificial Intelligence Lab., NASA Ames Research Center, Moffett Field, CA 94035, USA}
\affiliation{University of California Santa Cruz at NASA Ames Research Center, Moffett Field, CA 94035, USA}

\author{Bryan O'Gorman}
\affiliation{Quantum Artificial Intelligence Lab., NASA Ames Research Center, Moffett Field, CA 94035, USA}
\affiliation{SGT Inc., 7701 Greenbelt Rd, Suite 400, Greenbelt, MD 20770, USA}

\author{Joseph Fluegemann}
\affiliation{Quantum Artificial Intelligence Lab., NASA Ames Research Center, Moffett Field, CA 94035, USA}
\affiliation{San Jose State Research Foundation at
NASA Ames Research Center, Moffett Field, CA 94035, USA}

\author{Rupak Biswas}
\affiliation{Exploration Technology Directorate, NASA Ames Research Center, Moffett Field, CA 94035}

\author{Vadim N. Smelyanskiy}
\affiliation{Google, 150 Main St, Venice Beach, CA, 90291}

\date{\today}

\pacs{}
\begin{abstract}
Calibration of quantum computing technologies is essential to the effective utilization of their quantum resources. 
Specifically, the performance of quantum annealers is likely to be significantly impaired by noise in their programmable parameters, effectively misspecification of the computational problem to be solved, often resulting in spurious suboptimal solutions. 
We developed a strategy to determine and correct persistent, systematic biases between the actual values of the programmable parameters and their user-specified values. 
We applied the recalibration strategy to two D-Wave Two quantum annealers, one at NASA Ames Research Center in Moffett Field, California, and another at D-Wave Systems in Burnaby, Canada. 
We show that the recalibration procedure not only reduces the magnitudes of the biases in the programmable parameters but also enhances the performance of the device on a set of random benchmark instances.
\end{abstract}
 
\maketitle 

\section{Introduction}

Quantum annealing (QA) is a metaheuristic for solving combinatorial optimization problems \cite{kadowaki1998}. 
The recent introduction of QA hardware by D-Wave Systems~\cite{Bunyk_IEEE2014,johnson_quantum_2011} has invigorated theoretical and experimental research into the computational power and practical implementation challenges of the QA paradigm. 
Current research studies focus on both fundamental and applied aspects, including 
application to real-world problems ~\cite{PerdomoOrtiz2012_LPF, Gaitan2012, PerdomoOrtiz_EPJST2015, RieffelQIP2015, OGormanEPJST2015}, 
criteria for detecting quantum speedup~\cite{ronnow2014}, 
the computational role of quantum tunneling~\cite{boixo2015}, 
error-supression~\cite{pudenz2014},  
the relationship between classical simulated annealing and quantum annealing~\cite{boixo_NatCommun2013, boixo_evidence_2014,Shin_arXiv2014,  Albash_EPJST2015, MartinMayor_arXiv2015, Hen_arXiv2015,King_arXiv2015}, 
spin-glass perspectives on the hardness of computational problems~\cite{Katzgraber_PRX2015,Venturelli_arXiv2014}, and 
programming strategies that address intrinsic noise~\cite{PerdomoOrtiz_arXiv2015a,King_arXiv2014}.

The quantum annealers used for this study are of the second generation of D-Wave devices, also called D-Wave Two~\cite{Bunyk_IEEE2014}: one located at NASA Ames Research Center in Moffett Field, California, (``NASA device''), and another located at D-Wave Systems in Burnaby, Canada (``Burnaby device''). 
These consist of 64 unit cells of a previously characterized eight-qubit unit cell~\cite{harris2010,johnson_quantum_2011}. 
In the NASA and Burnaby devices, post-fabrication characterization determined that only 509 and 424 qubits, respectively, out of the 512 qubit arrays can be reliably used for computation. 
The array of coupled superconducting flux qubits is, effectively, an artificial Ising spin system with programmable spin-spin couplings and transverse magnetic fields. It is designed to solve instances of the following (NP-hard~\cite{barahona1982}) classical optimization problem: Given a set of local fields $\{h_i\}\subset\mathbb{R}$ and couplings $\{J_{ij}\}\subset\mathbb{R}$, find the assignment $\mathbf{s^*} = s^*_1 s^*_2 \cdots s^*_n$, that minimizes the objective function $E(\mathbf{s})$, 
\begin{equation}\label{eq:E-Ising}
E_{\mathrm{Ising}}(\mathbf{s})  = \sum_{1 \le i \le n} h_{i} s_i  + \sum_{1 \le i<j\le n} J_{ij} s_{i} s_{j},
\end{equation}
where $\abs{h_i} \le 2$, $\abs{J_{ij}} \le 1$, and $s_i \in \{\pm1\}$. 
Finding the optimal $\mathbf{s^*}$ is equivalent to finding the ground state of the corresponding quantum Ising Hamiltonian $H_{\mathrm{Ising}}  =  \sum_{i} h_{i}\sigma_{i}^{z}  + \sum_{i<j} J_{ij}\sigma_{i}^{z} \sigma_{j}^{z}$,
where $\sigma_{i}^{z}$ is Pauli $z$ operator acting on the $i$th spin. 
More details of QA can be found in Appendix~\ref{app:qa}.

Currently, D-Wave devices are only calibrated at the level of ensuring that the low-level control circuitry has its intended effect on the physical quantities like current, flux, etc., that it is meant to control~\cite{harris2010b,trevor}. Early research into the performance of D-Wave devices has indicated the presence of significant imprecision in the setting of the fields that define the problem to be solved, a consequentially significant impairment to the successful solution of the problem~\cite{boixo_evidence_2014,Venturelli_arXiv2014,King_arXiv2015,PerdomoOrtiz_arXiv2015a}. 
Recently, some work has used a phenomenological noise model of the fields $\{h_i\}$ and $\{J_{ij}\}$ in which the distributions of the deviations from the programmed values are given by Gaussians with means zero and standard deviations, respectively, of 0.05 and 0.035 (in units of the maximal $J_{ij}$), independently instantiated for each qubit and anneal and constant throughout the course of a given anneal.  
The parameters of the Gaussians were derived by adding in quadrature the variances of several known microscopic sources of noise~\cite{trevor}. 
This model has been used in an attempt to explain the failure rate of D-Wave devices as partly due to misspecification of the programmable values. 
There are many sources of noise in quantum annealers, each with a different effect and time scale, and we address here only one manifestation. 
The variances we cite for $\{h_i\}$ and $\{J_{ij}\}$ are in a sense incomparable to those just mentioned, and relevant only within the context of the experiments described below. 

The presence of systematic biases in quantum annealers has been reported elsewhere~\cite{King_arXiv2014}. 
The biases referred to there are fundamentally different in nature than the ones address here, in that the former are collective biases on the qubits of ferromagnetic chains that depend on the strength and topology of the couplings therein and are due to the noise specifically caused by those couplings, and they must be determined anew for each embedding topology used.
In this work, we present a methodology for determining, in parallel, the persistent, systematic biases in all of the individually available programmable parameters of a quantum annealer that requires a relatively small amount of total annealing time.
We show that correcting for these biases produces an increase in the quality of solutions found on a set of random benchmark instances.
The strategy presented here is the first proposal for a full-device recalibration at the user level, i.e. based only on the data from tailored instances and without access to the low-level control circuitry.
 \section{Determination of persistent biases}

Because actual quantum annealers operate at non-zero temperature, there exists some threshold for the values of the fields $\{h_i\}$ and couplings $\{J_{ij}\}$ below which thermal effects dominate the annealing process. When the strengths of the fields and couplings are set sufficiently small, the probabilities of the final states of the qubits are well described by a Boltzmann distribution. Roughly, the relevant energy scale is given by $kT$, where $k$ is Boltzmann's constant and $T$ is the temperature. (Henceforth, we will work in units in which $k=1$.) By running experiments in this regime, persistent biases in the values of the programmed fields can be uncovered. 

\subsection{$h$ biases}

In our model, $h_i=\hp_i + \hb_i$ is the effective value of the local field of qubit $i$, where $\hp_i$ is the user-programmed value and $\hb_i$ is the $\hp_i$-independent bias. In an ideal device, $\hb_i=0$.

Let ${p_i\big(\uparrow \big| \hp_i \big)}$ $\big[{p_i\big(\downarrow \big| \hp_i\big)}\big]$ be the probability of qubit $i$ being in the spin-up [spin-down] state at the end of an anneal with the programmed value $\hp_i$. A completely thermal model for this probability is given by
\begin{equation}
p_i\big(\uparrow \big| \hp_i\big) = \frac{e^{-\alpha_i\big(\hp_i, \hb_i, T_i\big)}}{e^{\alpha_i\big(\hp_i, \hb_i, T_i\big)} + e^{-\alpha_i\big(\hp_i, \hb_i, T_i\big)}},
\end{equation}
where $\alpha_i(\hp_i, \hb_i, T_i) \equiv h_i / T_i = \big(\hp_i + \hb_i\big)/T_i$ and $T_i$ is the temperature of qubit $i$.  This yields
\begin{equation}\label{eq:alpha-i}
 \alpha_i\big(\hp_i\big)\equiv \alpha_i\big(\hp_i, \hb_i, T_i\big) = \frac{1}{2}\ln\frac{1-p_i\big(\uparrow\big|\hp_i\big)}{p_i\big(\uparrow\big|\hp_i\big)}.
\end{equation}
(We assume that $h_i^{(b)}$ and $T_i$ are constant at least over the course of the experiment.)
More generally, we define the function ${\alpha(p)\equiv(1/2)\ln[(1-p)/p]}$. Once experimental values of $\alpha_i\big(\hp_i\big)$ are obtained for various values of $\hp_i$, the data are fit to obtain the estimates of $\hb_i$ and $T_i$. 

\subsection{$J$ biases}

The biases of the couplings between qubits can be determined in a similar fashion.
Let $J_{ij}=\Jp_{ij} + \Jb_{ij}$ be the effective value of the coupling between qubits $i$ and $j$ and $p_{ij}\big(\uparrow\uparrow\big|\hp_i, \hb_i, \hp_j, \hb_j, \Jp_{ij}, \Jb_{ij}\big)$ be the probability of qubits $i$ and $j$ both being in the spin-up state, where $\Jp_{ij}$ is the programmed value of the coupler and $\Jb_{ij}$ the bias, analogous to $\hp_i$ and $\hb_i$, respectively. 
Here, we set $\hp_i = \hp_j =0$, and write simply ${p_{ij}\big(\uparrow\uparrow\big| \Jp_{ij}\big) \equiv p_{ij}\big(\uparrow\uparrow\big|\hb_i, \hb_j, \Jp_{ij}, \Jb_{ij}\big)}$. 
Other probabilities ${p_{ij}\big(\cdot\cdot\big|\Jp_{ij}\big)}$, and combinations thereof such as ${p_{ij}\big(\uparrow\uparrow \lor \downarrow\downarrow \big| \Jp_{ij}\big) \equiv p_{ij} \big( \uparrow\uparrow \big| \Jp_{ij}\big) + p_{ij} \big(\downarrow\downarrow \big| \Jp_{ij}\big)}$, are analogously denoted. 

One approach to determine the bias $\Jb_{ij}$ is to naively assume that ${\hb_i=\hb_j=0}$, in which case the thermal distribution is modeled by 
\begin{equation}
p_{ij}\big(\uparrow\uparrow \lor \downarrow\downarrow\big|\Jp_{ij}\big) 
= \frac{2e^{-\alpha_{ij}\big(\Jp_{ij}\big)}}
{2e^{\alpha_{ij}\big(\Jp_{ij}\big)} + 2e^{-\alpha_{ij}\big(\Jp_{ij}\big)}}
\end{equation}
where $\alpha_{ij}\big(\Jp_{ij}\big) \equiv J_{ij} / T_{ij} = \big(\Jp_{ij} + \Jb_{ij}\big) / T_{ij}$.
For concision, we leave the dependence of $p_{ij}$ and $\alpha_{ij}$ on $h_i$, $h_j$, $\Jb_{ij}$, and $T_{ij}$ implicit.
Similarly to the case for $h_i$, this yields
\begin{equation}
\label{eq:approx-alpha-ij}
\alpha_{ij}\big(\Jp_{ij}\big) = \frac{1}{2}\ln\frac{1-p_{ij}(\uparrow\uparrow\lor\downarrow\downarrow)}{p_{ij}(\uparrow\uparrow\lor\downarrow\downarrow)}.
\end{equation}

A more accurate estimate for $\Jb_{ij}$ can be obtained by considering nonzero $\hb_i$ and $\hb_j$ (but still setting ${\hp_i=\hp_j=0}$). Let 
\begin{equation}
Z\equiv\sum_{s_i, s_j \in \{\pm 1\}} e^{(s_i\hb_i + s_j\hb_j + s_is_j(\Jp_{ij} + \Jb_{ij}))/T_{ij}}
\end{equation} be the partition function. Then
\begin{eqnarray}
\lefteqn{\frac{p_{ij}(\uparrow\uparrow)p_{ij}(\downarrow\downarrow)}{
p_{ij}(\uparrow\downarrow)p_{ij}(\downarrow\uparrow)}}\\ 
&=& \frac{e^{-(\hb_i + \hb_j +J_{ij})/T_{ij}}
e^{-(-\hb_i-\hb_j+J_{ij})/T_{ij}}/Z^2}
{e^{-(\hb_i-\hb_j-J_{ij})/T_{ij}}
e^{-(-\hb_i+\hb_j-J_{ij})/T_{ij}}/Z^2} \\
&=& e^{-4J_{ij}/T_{ij}},
\end{eqnarray}
or 
\begin{equation}
\label{eq:true-alpha-ij}
\alpha_{ij}(\Jp_{ij}, \Jb_{ij}, T_{ij}) = \frac{1}{4}\ln\frac{p_{ij}(\uparrow\downarrow)p_{ij}(\downarrow\uparrow)}{p_{ij}(\uparrow\uparrow)p_{ij}(\downarrow\downarrow)}.
\end{equation}
Note that the assumption that $\hb_i=\hb_j=0$ implies that $p_{ij}(\uparrow\uparrow)=p_{ij}(\downarrow\downarrow)$ and $p_{ij}(\uparrow\downarrow)=p_{ij}(\downarrow\uparrow)$, and that in this case~\eqref{eq:true-alpha-ij} reduces to~\eqref{eq:approx-alpha-ij}.

 \section{Biases determined in current D-Wave devices}
In this section and the next, we present preliminary results from the application of the above methods to actual hardware, namely two D-Wave quantum annealers: the NASA and Burnaby devices. Data collection was severely constrained by repeated downtimes and subsequent recalibrations of the device at NASA, and the presently ongoing installation of a newer chip, as well as limited access to the device at D-Wave Systems. This manuscript will be updated once more complete data has been collected.

\subsection{$h$ biases}
\label{sec:biases-determined-h}

To experimentally determine the biases $\{\hb_i\}$, we set all $\Jp_{ij}=0$ and initially assume the effect of nonzero $\{\Jb_{ij}\}$ to be negligible. 
We therefore ran the experiments for all qubits within a given device simultaneously, with the same value of $\hp_i=\hp$ for every working qubit.
Each value of $\hp$ was run 100 times, where each run consisted of 1,000 annealing cycles.
The probability $p_i^{(r)}\big(\uparrow\big)$ was calculated for each run $r$, and the median probability $\tilde{p}_i\big(\uparrow \big| \hp_i \big)$ taken over the 100 $\big\{p_i^{(r)} \big( \uparrow \big| \hp_i \big) \big\}$ calculated. 
From this, we define the ``median''\footnote{Technically, this is a slight abuse of terminology; while the median of $\big\{ \alpha \big( p_i^{(r)} \big) \big\}$ is almost the same as $\alpha\big(\tilde{p_i}\big)$ because the function $\alpha\big(p\big)$ is monotonic, the two quantities can differ slightly in the case of an even number of values.} $\tilde{\alpha}_i\big(\hp_i\big) = \alpha\big(\tilde{p}_i \big( \uparrow \big| \hp_i \big) \big)$.

\paragraph*{Calculation of the biases}
Since $\alpha_i  = \hp_i/T_i + h^{(b)}_i/T_i$, for each qubit, we fit a line to $\{(\hp_i, \tilde{\alpha}_i)\}$ by minimizing the quadratic loss; the resulting slope gives us an estimated inverse qubit temperature $\tilde{\beta}_i=1/\tilde{T}_i$, and from the intercept $\tilde{\alpha}_i^{(b)}$ we can determine the bias in several ways. 

One way is to simply use the fitted parameters as is: $\tilde{h}^{(b)}_i = \tilde{\alpha}_i^{(b)} \tilde{T}_i$. 
Experimental data, however, indicate that the estimates of the qubit ``temperatures'' calculated as above are not exactly that, but include in their calculation effects other than that due to true variation in temperature between the qubits. Some estimate of a uniform device temperature should therefore be used. 
(See App.~\ref{app:TivsTtilde} for more detail.)
In our experiments, we used two different quantities. 
The first is the ``mean temperature'', $\bar{T}F^{(h)}=\frac{1}{n}\sum_{i=1}^n\tilde{T}_i$, where $\tilde{T}_i=1/\tilde{\beta}_i$ and $n$ is the number of (working) qubits. 
The second is the ``median temperature'' $\tilde{T}^{(h)}$, called thus not because it is the median of $\{\tilde{T}_i\}$ but because it is calculated by taking the inverse of the slope of the line fit to the points $\{(\hp, \tilde{\alpha}^{(h)})\}$, where $\tilde{\alpha}^{(h)}\big(\hp\big)$ is defined as the median over the qubits of $\big\{\tilde{\alpha}_i\big(\hp\big)\big\}$. In practice, the quantities $\tilde{T}^{(h)}$ and $\bar{T}^{(h)}$ are effectively the same.

\begin{figure*}
\centering
\includegraphics[width=0.95\textwidth]{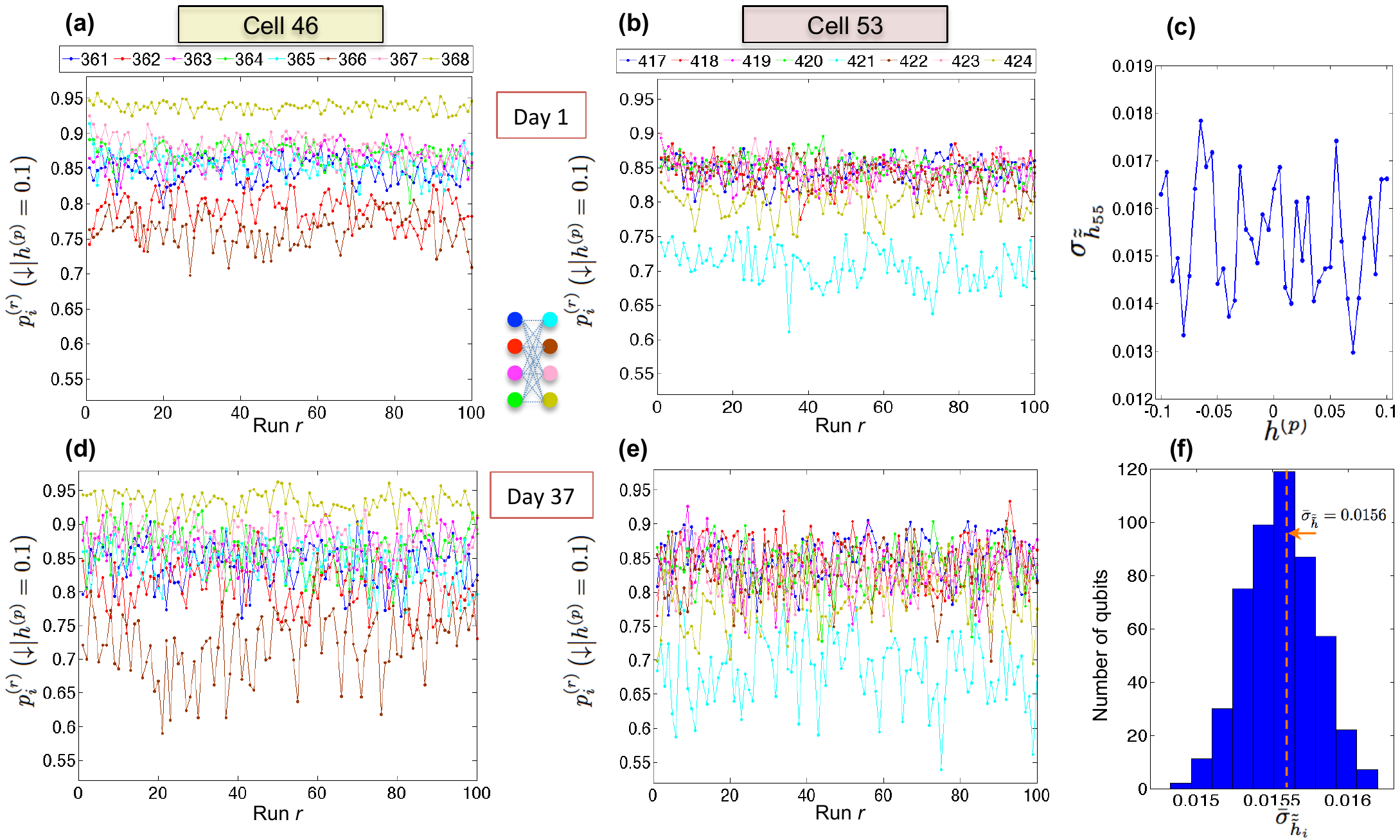}
\caption{
\textbf{Persistent systematic biases and inherent $h_i$ noise limit.}
(a,b,d,e) The probability $p_i^{(r)}\big(\downarrow\big|\hp=0.1\big)$ of qubit $i$ being spin-down in the 100 runs $\{r\}$ of 1000 reads each, for two experiments on the 16 qubits of two adjacent unit cells, done 36 days apart and in which the programmed value of all the qubits was $\hp=0.1$. 
Two important aspects are apparent that lay the foundation for the present work: there are qubits whose deviation from the rest is much greater than the noise level, and that this deviation persists for more than a month. As we show, these deviations in the probabilities are indicative of biases in the programmable parameters.
(c) For a single, typical qubit, 55, the standard deviation over 100 runs of the estimates $\left\{\bar{\tilde{h}}_{55}^{(r)}\right\}$ of $h_{55}$ versus the programmed $h^{(p)}$.
(f) The means $\big\{\bar{\sigma}_{\tilde{\tilde{h}}_i}\big\}$ over the programmed values $\big\{h^{(p)}\big\}$ of the standard deviations $\big\{\sigma_{\tilde{\tilde{h}}_i}\big\}$, as in (c).
The values of $\tilde{\tilde{h}}_i$ in (c,f) were calculated using the mean temperature $\tilde{T}^{(h)}$.
}
\label{fig:noise}
\end{figure*}

\paragraph*{Inherent noise limit}
Figures ~\ref{fig:noise}(a,b,d,e) shows the probability $p_i^{(r)}=p_iF^{(r)}\big(\downarrow|\hp_i=0.1\big)$ for each of the qubits in two different unit cells  and for each of the 100 runs. 
From this we see two fundamental phenomena: the run-to-run fluctuation of the probability $p_i^{(r)}$, and the presence of some qubits for which the typical probability is well separated from the others, to a degree greater than the inherent variability. 
The latter indicates the presence of the qubit-specific biases that we hope to determine and correct, while the former provides a limit to the precision with which we can control the effective $h_i$. 
We consider a calibration procedure successful if post hoc the qubits are unbiased to within this experiment limit.
To compare the variability of $\big\{p_i\big(\hp\big)\big\}$ over different values of $\hp$, for each qubit $i$ and value of $\hp$ we compute the 100 values of $\tilde{\tilde{h}}_i^{(r)}=\alpha\big(p_i^{(r)} \big( \hp \big) \big)\tilde{T}^{(h)}$ from the 100 runs. 
Figure ~\ref{fig:noise}(e) shows the standard deviation $\sigma_{\tilde{\tilde{h}}_{55}^{(r)}}$ of $\big\{\tilde{\tilde{h}}_{55}^{(r)}\big\}$ over the 100 runs; the plot is typical. 
Importantly, the standard deviation $\sigma_{\tilde{\tilde{h}}_i^{(r)}}$ is seemingly independent of the programmed $\hp_i$ for the values of $\hp$ considered. 
To characterize the typical variance of $h_i$, for each qubit the mean $\bar{\sigma}_{\tilde{\tilde{h}}_i}$ of standard deviations $\big\{\sigma_{\tilde{\tilde{h}}_i^{(r)}}\big\}$  was taken over the values of $\hp$. 
Figure~\ref{fig:noise}(f) shows a histogram of this mean standard deviation over all of the qubits. 
The distribution is quite tight, with an average of $\bar{\bar{\sigma}}_{\tilde{\tilde{h}}} = 0.0156$.
This run-to-run variation in the estimate of $h_i$ for a given $\hp$ ultimately leads to a limit on the precision with which we can estimate $\hb_i$, though quantitatively the exact limit depends on the number of runs, number of reads per run, and values and number of $\hp$s examined.

\begin{figure*}
\centering
\includegraphics[width=0.95\textwidth]{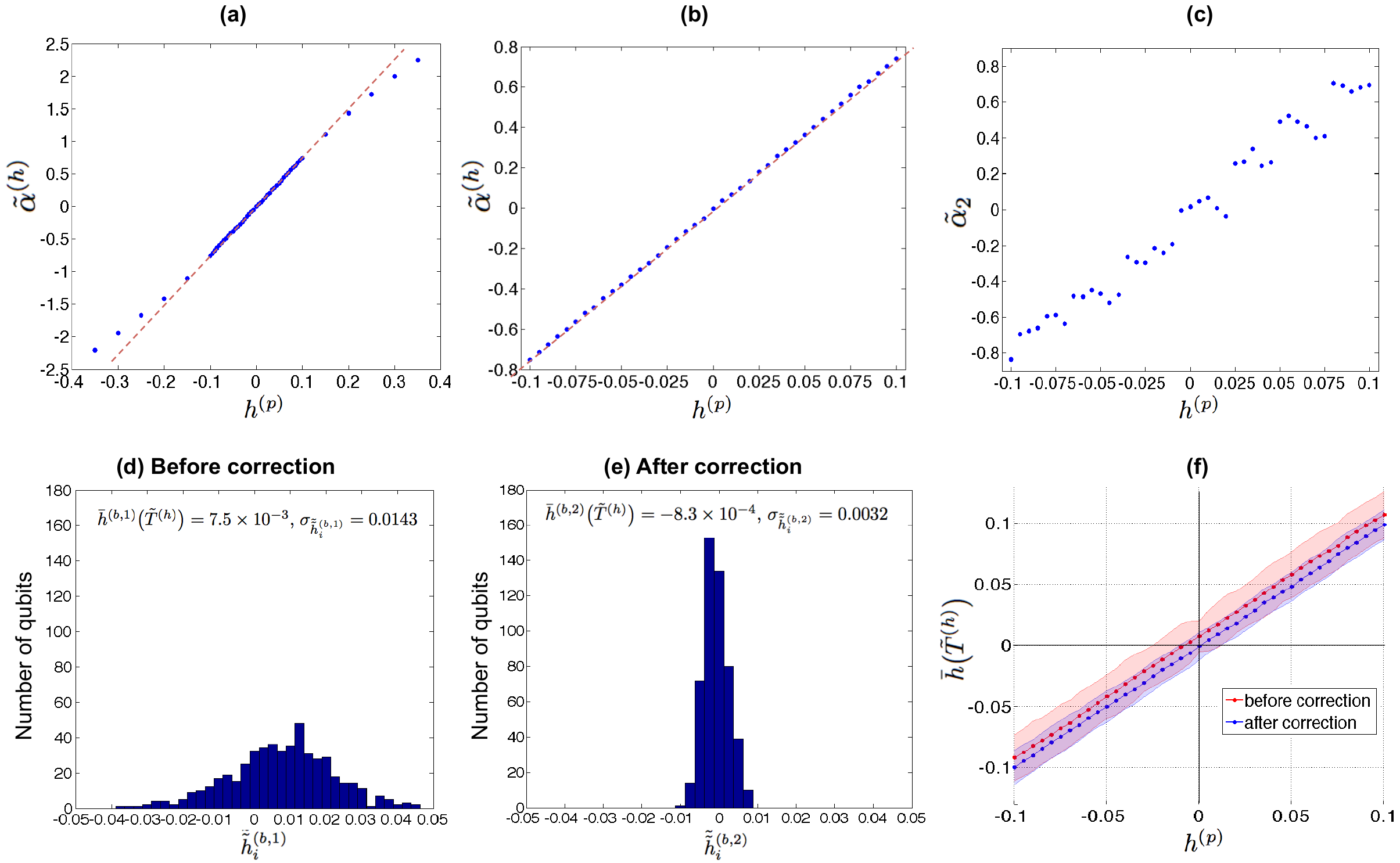}
\caption{
\textbf{Detection and correction of systematic biases.}
(a) From a single experiment without correction, the median quantity $\tilde{\alpha}^{(h)}\left(\hp\right)$, over the qubits, of the quantities $\left\{\tilde{\alpha}_{i}\left(\hp\right)\right\}$ for various values of $\hp$ in [-0.35,0.35].
(b) Same as (a), but only using values of $\hp$ in [-0.1,0.1]. Note the tightness of the fit to a line, indicating the validity of the thermal model.
(c) From the same experiment as in (b), the quantity $\tilde{\alpha}_2\big(\hp\big)$ for a single, typical qubit, 2. Note the step function resulting from the limited precision of the digital-to-analog converter used to control the field.
(d,e) The estimated biases $\big\{\tilde{\tilde{h}}_i^{(b,k)}\big\}$ from two experiments $k=1,2$. The first is without correction; the second was corrected using the biases estimated from the first. Note that the distribution significantly more narrowed and centered near zero after correction.
(f) The average over the qubits of the quantites $\big\{\tilde{\tilde{h}}_i\big\}$ for different values of the programmed $\hp$, with error bands given by the standard deviation. As for (d,e), note the narrowing and centering of the distributions.
}
\label{fig:hscans}
\end{figure*}

Figures ~\ref{fig:hscans}(a,b) shows two windows of $\tilde{\alpha}^{(h)}\big(\hp\big)$ for one experiment on the device at NASA Ames. Figure ~\ref{fig:hscans}(b) shows the range $\hp\in [-0.1,0.1]$ used in the calculation of the biases. There, the linearity of $\tilde{\alpha}^{(h)}$, and thus the accuracy of the thermal model, is evident; this is typical of all the experiments reported. Figure ~\ref{fig:hscans}(a) shows a wider window $\hp\in [-0.35,0.35]$, where the nonlinearity outside of the former range is evident, indicating the failure of the thermal model for larger magnitudes of $\hp$, where annealing dynamics start to dominate thermal dynamics. Figure ~\ref{fig:hscans}(c) shows $\tilde{\alpha}_2\big(\hp\big)$, revealing the limited resolution ~0.025 of the digital-to-analog converters (DACs) used to implement $h_2$. Such stepping behavior is typical for all of the qubits.

\paragraph*{Narrowing of the bias distribution}

To show the correctability of the persistent biases, we run the experiment described above repeatedly, each time attempting to correct the biases using estimates thereof from the prior iteration. 
Let $\hbi{k}_i$ $\big[\tilde{T}^{(h,k)}\big]$ be the experimentally determined value of the bias $\hb_i$ $\big[\text{median temperature }\tilde{T}^{(h)}\big]$ based on the $k$-iteration; by convention, set $\hbi{0}_i=0$. Let $\hpi{0}$ be the desired programmed value. In the $k$-th iteration, we set 
\begin{equation}
\label{eq:hpik}
\hp_i=\hpi{k}_i=\hpi{0}-\sum_{k'=1}^{k}\hbi{k'}_i.
\end{equation}

Figures~\ref{fig:hscans}(d-e) show the distribution of $\big\{\tilde{h}^{(b,k)}_i \big\} \equiv \big\{\tilde{\alpha}^{(b,k)}_i \tilde{T}^{(h,k)}\big\}$ for the first and second iterations, i.e. before and after the recalibration procedure. 
The narrowing in the distribution is a clear indication that the procedure is working to remove the extreme biases. 

Further support for the success of recalibration is provided by looking at the distribution over the qubits of the success probabilities for all values of $\hp$.
Figure~\ref{fig:hscans}(f) shows a uniform reduction of the variance over the qubits in the values of $\tilde{h}$. 
The points are the mean $\bar{h}$ over the qubits of $\big\{\tilde{h}_i \big\}$ for the corresponding $\hp$, and the shaded region indicates the standard deviaton. 
The narrowing of the distribution is clear evidence that the recalibration procedure not only narrows the distribution of the biases [Fig.~\ref{fig:hscans}(d,e)], as reflected in the shift of the mean, but also reduces the variance for all values of $\hp$. 

\subsection{$J$ biases}

\paragraph*{Calculation of the $J$ biases}
In the data presented here, the $J$ biases were determined using Eq. ~\ref{eq:approx-alpha-ij}. 
The programmable value of the local fields $\big\{\hp_i\big\}$ were uniformly set to zero. 
For a given value of $\Jp$, the experiment was run in six batches. 
In each batch, the programmed coupling $\Jp_{ij}$ was uniformly set to $\Jp$ for a each coupler of a pairwise disjoint subset of all of the couplers, and for each of the rest $\Jp_{ij}$ was set to zero. 
Over the six batches, each coupling $\Jp_{ij}$ was set to $\Jp$ exactly once.
As for the $h$ biases, each value of $\Jp$ was run 100 times (for each coupler), with each run consisting of 1,000 annealing cycles. 
The median probability $\tilde{p}_{ij}\big(\uparrow\uparrow \lor \downarrow\downarrow \big| \Jp_{ij}\big)$ was taken over the 100 $\big\{p^{(r)}_{ij}\big\}$ calculated from the runs, from which we calculate the ``median'' $\tilde{\alpha}_{ij}\big(\Jp_{ij}\big) = \alpha\big(\tilde{p}_{ij}\big(\uparrow\uparrow \lor \downarrow\downarrow \big| \Jp_{ij}\big)\big)$. 
For each coupler, a line was fit to $\big\{\big(Jp_{ij}, \tilde{\alpha}_{ij}\big)\big\}$, yielding a slope $\tilde{\beta}_{ij}=1/\tilde{T}_{ij}$ and an intercept $\tilde{\alpha}_{ij}^{(b)}$. 
As for $h$,  we define the mean temperature $\bar{T}^{(J)}=\frac{1}{m}\sum_{\{i,j\}}\tilde{T}_{ij}$, where $m$ is the number of couplers.

\begin{figure*}[t!]
\centering
\includegraphics[width=0.95\textwidth]{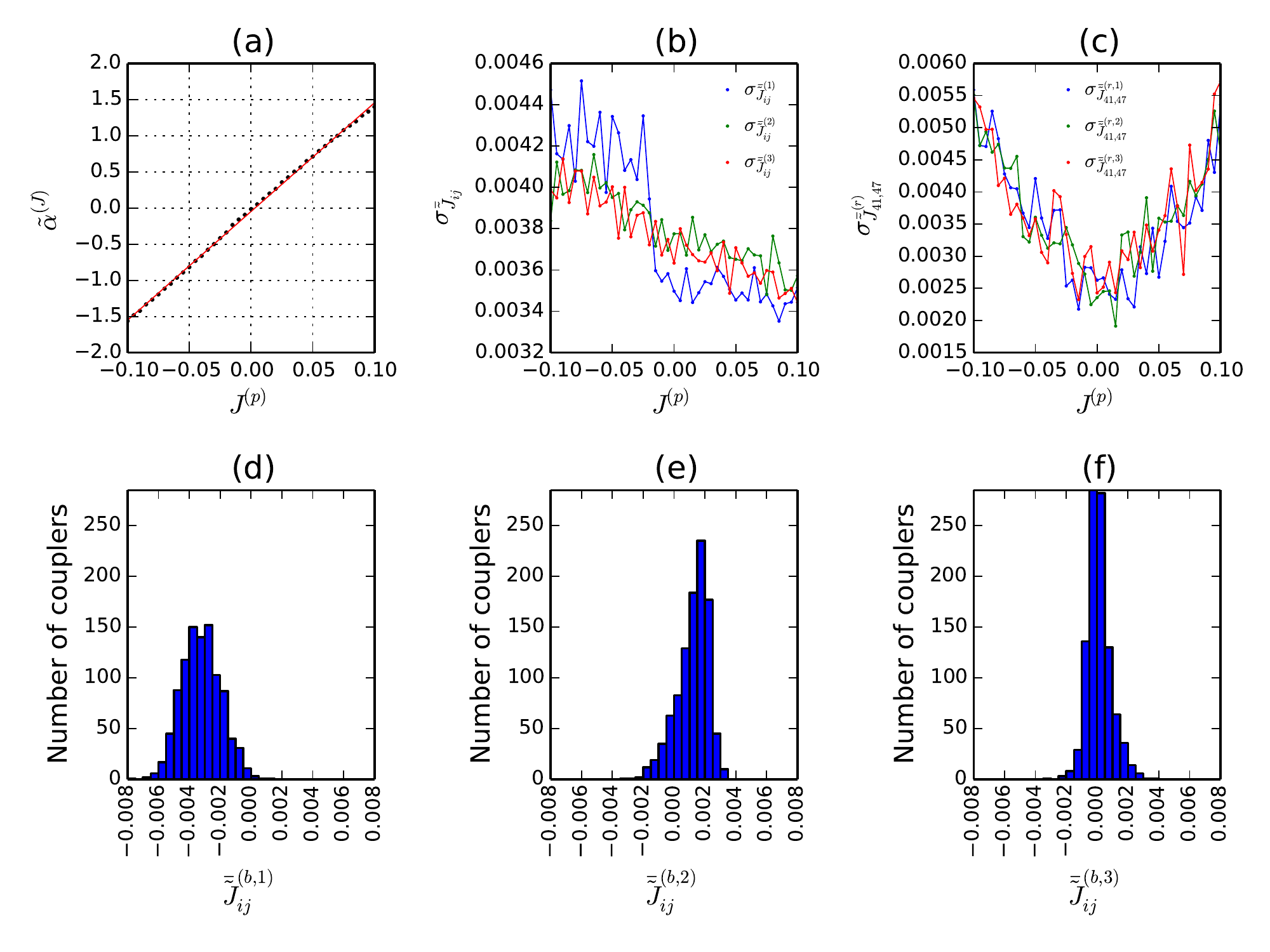}
\caption{\textbf{$J$ biases in the Burnaby device.} 
Data from a series of three experiments ($k=1,2,3$) in which for each experiment the sums of biases estimated in the previous ones are subtracted from the original programmed values, using Eq.~\ref{eq:Jpik}. 
The first experiment is without any correction, and the second and third use increasingly accurate corrections. 
All quantities are calculated using the mean $\bar{T}^{(J,k)}$ of the qubit temperatures, calculated indepently in each experiment.
(a) From only the first experiment, the median quantity $\tilde{\alpha}^{(J)}\left(\Jp\right)$, over the couplers, of the quantities $\left\{\tilde{\alpha}_{ij}\left(\Jp\right)\right\}$ for evenly spaced values of $\Jp$ in [-0.1,0.1].
(b) The standard deviation over the couplers of the estimated $\bar{\tilde{J}}_{ij}^{(k)}$ at each value of the original $J^{(p,0)}$.
(c) For a single, typical coupler, (41, 47), the standard deviation over 100 runs of the estimates $\left\{\bar{\tilde{J}}_{41,47}^{(r)}\right\}$ of $J_{41,47}$ versus the original $J^{(p,0)}$.
(d-f) Residual biases $\bar{\tilde{J}}_{ij}^{(b,k)}$ estimated from each of the experiments.
}
\label{fig:combined-J-plots}
\end{figure*}

Figure~\ref{fig:combined-J-plots}(a) shows the median quantity $\tilde{\alpha}^{(J)}$ for evenly spaced values of $\Jp$ in [-0.1, 0.1], as well as a line fit thereto. 
The closeness of the fit of the line confirms the accuracy of the thermal model as for $\tilde{\alpha}^{(h)}$. 

Define $\Jbi{k}_{ij}$ to be the estimate $\tilde{\alpha}_{ij}^{(b,k)}\bar{T}^{(J,k)}$ for $\Jb_{ij}$ using the data from the $k$-th iteration, with $\Jbi{0}_{ij}=0$ by convention. In the $k$th iteration, we set the programmed values of the couplers to 
\begin{equation}
\label{eq:Jpik}
\Jp_{ij}=\Jpi{k}_{ij}=\Jpi{0}-\sum_{k'=1}^{k}\Jbi{k'}_{ij},
\end{equation}
i.e. by subtracting the sums of the residual biases from the prior iterations from the desired values. 
Figures~\ref{fig:combined-J-plots}(d-f) show the narrowing of the distribution of residual $J$ biases with correction. 

Unlike the case for the $h$ biases, for which data indicate that the distribution is essentially converged after a single iteration, here we see two new phenomena. First, the distribution continues to narrow between the second and third iterations. Second, the distribution of the residual biases from the second iteration, while narrower than that from the first, is not centered around zero. We believe this is due to overcorrection; that is, the estimates of the biases from the first iteration have a high degree of uncertainty, and so simply subtracting their values from the intended value introduces some amount of bias itself. This is consistent with the overall small magnitudes of the $J$ biases relative to those of the $h$ biases, especially as compared to the corresponding noise levels. This overcorrection can be mitigated by weighting the correction in a way that accounts for the uncertainty in the estimate using Bayesian reasoning.

Figure~\ref{fig:combined-J-plots}(b), analogous to Fig.~\ref{fig:TivsTtilde}(b), shows the standard deviation $\sigma_{\bar{\tilde{J}}_i}$ over the qubits of the estimates $\big\{\bar{\tilde{J}}_{ij}^{(k)}\big\}$ of $\big\{J_{ij}\big\}$ for different values of $\Jp$, for each of the three iterations of experiments as described above. The estimates $\big\{\bar{\tilde{J}}_{ij}^{(k)}\big\}$ were calculated independently for each iteration $k$ using its mean temperature $\bar{T}^{(J,k)}$. There is a significant change from the first iteration of corrections, but then the standard deviation remains approximately the same after the second iteration. Considering the average over the values of $\Jp$, the overall variance is about the same before and after correction, yet is much more uniform after correction, which we consider beneficial. 

Figure~\ref{fig:combined-J-plots}(c), analogous to Fig.~\ref{fig:noise}(c), shows the standard deviation $\sigma_{\bar{\tilde{J}}^{(r,k)}_{41,47}}$ of the estimates $\big\{\bar{\tilde{J}}^{(k)}_{41,47}\big\} = \big\{\tilde{\alpha}^{(k)}_{41,47}\bar{T}^{(J,k)}\big\}$. The correction seems to have no effect, and the there is a consistent increase in the variance with increasing magnitude of the programmed value $\Jp$. A similar phenomenon occurred in the analogous $h$ data.
 \section{Effect of correction on performance}
Ultimately, the goal of calibration is to optimize the performance of a quantum annealer on problems of computational interest. It is not clear a priori that the biases present in one- and two-qubit experiments are the same as those present in anneals involving hundreds of qubits. Even if they were, their estimation would be of no practical value unless their correction improves performance.  
To address this, we tested the effect of correcting the $h$ biases on the performance of the quantum annealer at NASA Ames, using the same parameterized random ensemble of instances used in a previous study benchmarking a D-Wave quantum annealer~\cite{ronnow2014}. 
To generate a single instance with range $r$, the local fields $\{h_i\}$ were uniformly set to zero, and each available $J_{ij}$ was independently and uniformly selected from $\{-r, -r+1, \ldots, -1, 0, 1, \ldots, r\}$. The resulting instance was then scaled by the overall factor $0.9/r$ so that the largest magnitude $|J_{ij}|$ was 0.9. 
(This was necessary, rather than scaling to 1 as in previous studies, to allow for consistency with future experiments in which the $J$ biases are corrected.)
100 such instances were generated and run twice with 1,000 annealing cycles for each of the same (uniformly randomly generated) 10 gauges. 
In all runs, $\{J_{ij}\}$ were programmed as in the instances. 
For the first set of runs, which we call ``uncorrected'', the $\{h_i\}$ were also programmed as in the instances, i.e. to zero. 
For the other set, which we call ``$h$-corrected'', the local fields were programmed to the inverse of the biases computed via experiments as in Eq.~\ref{eq:hpik}.

For each instance, the uncorrected and corrected results were compared using two methods, a ``greedy'' one and the elite mean. The greedy comparison is as follows: the energies of all states returned were computed, and those for all gauges were grouped together. 
Whichever method (uncorrected or corrected) returned the lower minimum energy was deemed to have performed better. 
If the minimum energies were the same, the tie was broken by the number of times that energy was returned. 
If this number was the same, the method with the second-lowest energy was deemed to have performed better, with ties broken by the number of times the second-lowest energy was returned, and so on. 

The ``elite mean'' score function~\cite{PerdomoOrtiz_arXiv2015a}, a quantity previously introduced to allow comparison of the performance of different programming parameters in quantum annealers when the success probabilities are too low (and thus noisy), is defined as the mean energy of the ``elite'' states, i.e. those with the lowest energies. 
The elite mean is parameterized by the fraction of energies over which to take the mean; here we use 2\%. 
The results of the comparison are summarized in Table~\ref{tab:benchmarks-results}, showing the proportion of instances, for each $r$, for which the correction improved the performance, using each of the two comparison methods described above. For ${r=1}$ and ${r=2}$, there were 6 and 2 instances, respectively, for which the elite mean comparison was tied, all but one due to success probabilities greater than 2\% for both the corrected and uncorrected experiments.
The data set is too small to imply definite conclusions, but does indicate that correcting for the $h$ biases, even using data from one experiment, improves performance according to reasonable metrics.

\begin{table}[h]
\caption{
\textbf{Comparison of performance with and with $h$-correction on benchmarks.}
}
The probability that correcting for $h$ biases (using data from a single experiment) improved performance on 100 random instances from an ensemble parameterized by the range of values $r$. Performance was compared according to two metrics: greedy comparison of the energies and degeneracies, and comparison of the elite mean score function.
\label{tab:benchmarks-results}
\begin{ruledtabular}
\begin{tabular}{c|c|c|c|c|c}
Range $r_J$ & 1 & 2 & 4 & 8 & 16 \\ \hline
Greedy & 0.58 & 0.63 & 0.59 & 0.68 & 0.53 \\ \hline
Elite mean &  0.65 & 0.65 & 0.73 & 0.72 & 0.67 \\
\end{tabular}
\end{ruledtabular}
\end{table}
  \section{Conclusions}

Imprecision in the programmable parameters of a quantum annealer can significantly degrade its performance. Some amount of imprecision is unavoidable in an analog device, but we have shown how to correct that part of the deviation from the desired parameters that is persistent and systematic. In this work, we present a methodology for determining, in parallel, the persistent, systematic biases in all of the available programmable parameters $\big\{h_i, J_{ij}\big\}$ of quantum annealer that requires a relatively small amount of total annealing time and whose results remain valid for weeks on current devices.
The strategy presented here exploits the finite temperature of quantum annealers by using values of the programmable parameters such that $h_i, J_{ij} \lesssim k T$;
in this regime, the results of one- and two-qubit experiments are well described by a Boltzmann distribution, from a model of which the biases of interest can be extracted.

We applied our method to two D-Wave devices, one located at NASA Ames Research Center in Moffett Field, California, and one located at D-Wave Systems in Burnaby, Canada. 
The application of the method to both devices lead to more uniform statistics of one- and two-qubit experiments. In a small test of standard benchmark instances using all of the available qubits and couplers of the NASA device, the method improved performance.
Further studies on both the devices used here, as well as the only other available device (``LM-USC''), located at the University of Southern California-Lockheed Martin Quantum Computation Center in Marina del Rey, California are currently in progress.

Data collection was significantly impaired by recurrent maintenance on the NASA device and the recency of our access to the Burnaby and LM-USC device. 
More complete data from all three devices, currently being collected, will provide a stronger characterization of the noise, especially the effect of the programmed values of neighboring fields $\big\{h_i^{(p)}, J_{ij}^{(p)}\big\}$ on the effective value of a given $h_i$ or $J_{ij}$, as well as enable the analysis of extensions of the methods. 

The estimates of the biases can likely be made more accurate, as judged by their effect of their correction on subsequent estimations thereof and on performance on benchmark instances, given the same amount of annealing time. 
First, preliminary data indicates that calculating the $J$ biases without the assumption of no $h$ biases (i.e. using Eq.~\ref{eq:true-alpha-ij} as opposed to Eq.~ref{eq:approx-alpha-ij}) leads to more consistent estimates. 
Second, disentangling the mutual effect of the $h$ and $J$ biases on each other by alternating between iterations of the iterations of $h$ and $J$ experiments (as opposed to doing each alone as reported) will likely lead to more accurate estimates of each individually. 
Lastly, the risk of overcorrection can be mitigated by weighting the correction by the degree of certainty of the estimate of the bias to be corrected.

Although we focused initially on a standard random ensemble of Ising instances for benchmarking the performance of quantum annealers, the effect of correcting biases should be greatest on instances whose ground states are most sensitive to misspecification of the programmable parameters. 
On average, this sensitivity increases with the range $r$ of the random ensemble, because the values of the parameters are scaled to fit within the finite physical range while the biases remain at the same absolute scale. At a large enough range $r$, however, even correction of the biases is not enough, and inherent fluctuations lead to almost zero success probabilities. 
This would explain the possible pattern seen in the elite mean comparison (Table~\ref{tab:benchmarks-results}) that the advantage of correction peaks seems to peak at the level of $r$ considered to correspond to the precision limit of the device. 
(That such a pattern is not as apparent in the greedy comparison is easily explained by natural noisiness of that comparison method, especially for instances with extremely low success probability as was the case here.)

There is reason to suspect that correction will also have a beneficial effect in reducing the effect of gauge selection on success probability. 
While there are other suspected reasons for the effect of gauge selection (which would be non-existent in an ideal device), that biases could be one factor is plausible. 
The effect of gauge selection is significant, sometimes leading to an orders-of-magnitude difference in the success probabilities, and so this is a promising avenue for bias correction.

Importantly, while the $J$ biases determined here are in general smaller than the $h$ biases, numerical studies indicate that often instances are more sensitive to misspecification in the $J$ parameters than in the $h$ parameters\cite{zhu2015}.

The methods presented here complement a growing suite of tools for optimal programming of quantum annealers~\cite{PerdomoOrtiz_arXiv2015a,King_arXiv2014}, tuning the performance thereof to cope with the intrinsic noise in current and future physical implementations. 
 \section*{Acknowledgements}
We thank D-Wave Systems for the use of its device. 
This work was supported in part by the Intelligence Advanced Research Projects Activity (IARPA), the Office of the Director of National Intelligence (ODNI), via IAA 145483, and by the Air Force Research Laboratory (AFRL) Information Directorate under grant F4HBKC4162G001. 
The views and conclusions contained herein are those of the authors and should not be interpreted as necessarily representing the official policies or endorsements, either expressed or implied, of ODNI, IARPA, AFRL, or the U.S. Government. 
The U.S. Government is authorized to reproduce and distribute reprints for Governmental purposes notwithstanding any copyright annotation thereon. Institutional support was provided by the NASA Advanced Exploration Systems program and NASA Ames Research Center. The authors would like to thank T. Lanting, J. Realpe-Gomez, E. Rieffel, K. Kechedzhi, and D. Venturelli for useful discussions. The authors thank D-wave Systems, Inc for remote access to their 424-qubit device located in Burnaby, Canada.
 
\clearpage
\appendix

\section{Quantum annealing}\label{app:qa}

\subsection{Computational problem}
QA is designed to solve the classical problem of finding the ground state of an Ising system, i.e. finding an assignment of a set of $n$ classical spins $\mathbf s=\big(s_i\big)_{i=1}^n\in \{\pm 1\}^n$ that minimizes a given energy function.  
Any such function of spins has a unique representation as a sum of monomials of groups of spins, and generally the order (locality) of such monomials  is restricted to some constant. 
For any constant locality, a polynomial number of ancillary qubits can be introduced and a quadratic (2-local) energy function defined on the original and ancilla qubits  such that the  ground state of the original energy function can be directly inferred from the ground state of the 2-local energy function. For that reason, we restrict our attention to only 2-local energy functions of the form

\begin{equation}
\label{eq:ising}
E(\mathbf s) = \sum_{1\leq i \leq n} h_i s_i  + \sum_{1\leq i < j \leq n}J_{ij}s_is_J.
\end{equation}

The Ising problem is completely equivalent to the \textsc{Polynomial Unconstrained Binary Optimization} (\textsc{PUBO}) problem, defined as the minimization of pseudo-Boolean functions, i.e. real-valued polynomials of $\{0,1\}$-valued bits, through a linear relation between spins and bits. The restriction of \textsc{PUBO} to 2-local function is known as \textsc{Quadratic Unconstrained Binary Optimization} (\textsc{QUBO}).
Much of the literature on mapping arbitrary problems to quantum annealers and reducing the locality of energy functions uses the \textsc{PUBO} formalism, but we refer here only to the Ising formalism because it is more closely related to the physical implementation. The Ising problem is NP-hard \cite{barahona1982}.

In practice, physical constraints on the location of qubits and their couplers disallow such arbitrary connectivity, with an interaction between any pair of qubits. A further step known as embedding allows for hardware with restricted connectivity to solve problems of the form in Eq.~\ref{eq:ising} by mapping each logical qubit to a connected set of physical qubits and introducing strong ferromagnetic couplings between them to make them act as one.

\subsection{Physical process}
Quantum Annealing (QA) is a metaheuristic for solving combinatorial optimization problems \cite{kadowaki_quantum_1998}. The main procedure is this: a final Hamiltonian is constructed for some set of qubits such that its ground state encodes the optimal solution to the desired problem, those qubits are prepared in the ground state of an initial Hamiltonian, and then over the course of the annealing process the initial Hamiltonian is continuously transformed into the final Hamiltonian.
While there is ambiguity regarding the exact definition of the term ``QA'' and its differences with its closely related spin-off, adiabatic quantum computation\cite{Farhi2001}, we define it here to allow for non-adiabiticity in the annealing process and finite temperature, on the latter of which our methods are based.

To solve a computational problem, a Hamiltonian is constructed whose ground state encodes the optimal solution of the given problem,
\begin{equation}
H_{\text{final}} = \sum_{1\leq i \leq n} h_i\sigma_i^{(z)} + \sum_{1\leq i < j \leq n}J_{ij}\sigma_i^{(z)}\sigma_j^{(z)},
\end{equation}
i.e. the quantum analog of~\eqref{eq:ising}. The system is initialized in the easily-prepared ground state of another Hamiltonian, usually
\begin{equation}
H_{\text{init}} = \sum_{1\leq i \leq n} \sigma_i^{(x)},
\end{equation}
whose ground state is a uniform superposition of all $2^n$ computational basis states.
Then the Hamiltonian is slowly changed from the former to the latter; explicitly, 
\begin{equation}
H(t) = A(t) H_{\text{init}} + B(t) H_{\text{final}},
\end{equation}
where $A(t), B(t) \geq 0$ define the ``annealing profile'' and are such that $A(0), B(t_{\text{anneal}})>>0$ and $A(T)=B(0)=0$, where $t_{\text{anneal}}$ is the annealing time. 

\subsection{D-Wave devices}
D-Wave devices consist of an array of coupled superconducting flux qubits that is effectively a Ising spin system with programmable spin-spin couplings and local fields, longitudinal and transverse. 
The qubits are arranged in a so-called Chimera topology consisting of a square lattice array of bipartite unit cells.
The two devices used in this work consist of 512 nominal qubits in an 8-by-8 array of 8-qubit unit cells. 
Post-fabrication testing indicated that only 509 and 424 qubits were usable in the NASA and D-Wave devices, respectively. 
In both devices, the programmed values of $\{h_i\}$ and $\{J_{ij}\}$ are specified by unitless parameters in $[-2, 2]$ and $[-1, 1]$, respectively, where $1.0$ corresponds to an actual energy of $3.2$ GHz.
Further details of the hardware can be found in \cite{harris2010a}.
Thoughout this paper, we use the standard 1-based indexing for the qubits of a D-Wave device with 512 qubits, as in, e.g., Fig. 7(b) of~\cite{pudenz2014}. (While the set of functional qubits differs between actual devices, the convention is to still assign the indices to all qubits and to disregard the broken ones.)

\begin{widetext}

\clearpage
 \section{Persistency of the biases}
\label{sec:persistency}

Fig.~\ref{fig:persistency} shows the correlation of the biases $\big\{\tilde{\tilde{h}}^{(b,1)}_i\big\}$ as estimated from single experiments repeated at different times. 
Each data point corresponds to the bias in each of the 509 qubit of the NASA device. 
All experiments were performed as described in Sec.~\ref{sec:biases-determined-h}, with 100 runs of 1000 annealing cycles each for 41 evenly spaced values of $\hp$ in [-0.1,0.1]. 
Notice the strong correlation for intervals greater than one month. 
The experimentally determined median temperature $\tilde{T}^{(h)}$ can also be considered constant, with values within 1\% from the average of these four realizations: $\tilde{T}^{(h)}  = 19.0, 19.2, 19.2,$ and $19.2$ mK, for the experiments on 10-09-2014 at 14:00, on 10-09-2014 at 16:30, on 10-24-2014 at 11:15, and on 11-12-2014 at 7:45, respectively.

\begin{figure*}[h!]
\centering
\includegraphics[width=0.80\textwidth]{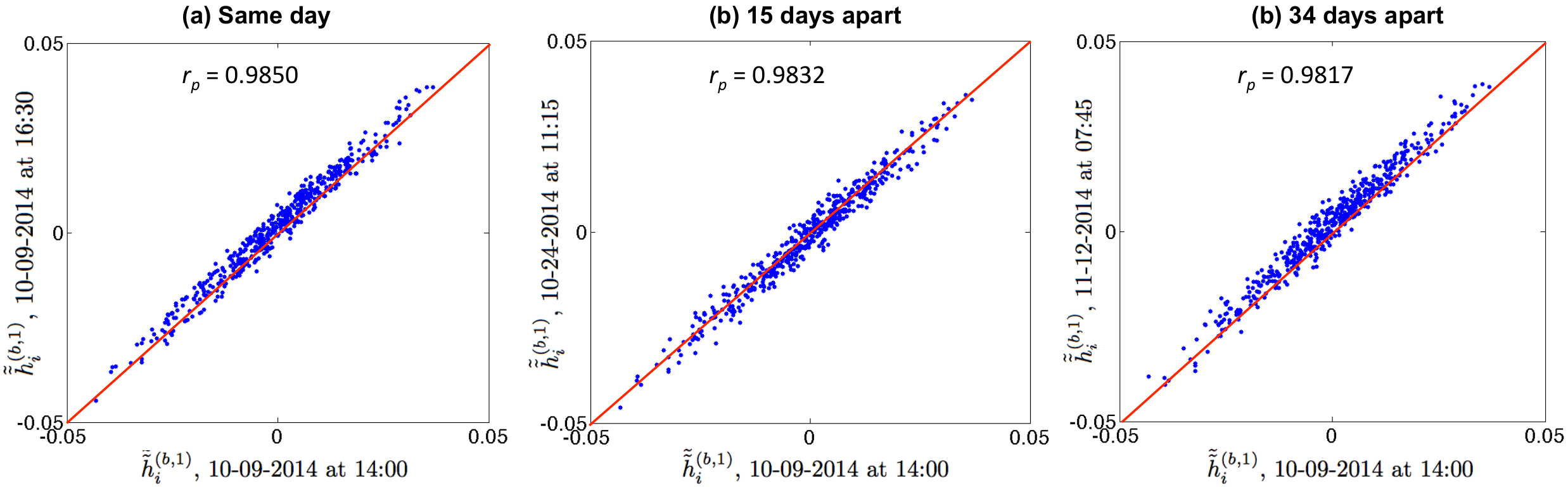}
\caption{\textbf{Persistency of systematic biases for time intervals.}}
\label{fig:persistency}
\end{figure*}

 \section{Qubit temperatures}
\label{app:TivsTtilde}
Figure~ \ref{fig:TivsTtilde}(a) shows the distribution of estimated ``qubit temperatures" $\{\tilde{T}_i\}$. 
Recall that each qubit temperature $T_i$ extracted a linear fits of the points $\{(\hp_i,\tilde{\alpha}_i)\}$, e.g. Fig.~\ref{fig:hscans}(c). 
That the variance of the temperatures of the qubits is as large as indicated here seems unlikely. 
For that reason, a more physically plausible ``device temperature'' was estimated, i.e. $\tilde{T}$ or $\bar{T}$, as described in Sec.~\ref{sec:biases-determined-h}. 
Data indicate that the variance is partly systematic, i.e. not due to the fluctuation of the experimental values on which the estimates is based. 
We suspsect that susceptibility~\cite{Boixo_arXiv2015} and cross-talk may play a role here.

Figure \ref{fig:TivsTtilde}(b) shows, for three related experiments on the NASA device, the standard deviation $\sigma_{\tilde{\tilde{h}}_i}$ over the qubits of the estimated value $\big\{\tilde{\tilde{h}}_i\big\}$ for different values of $\hp$. One experiment was without any correction; the other two used corrections calculated in two different ways, one using the qubit temperatures $\big\{\tilde{T}_i\big\}$ and the other the median temperature $\tilde{T}^{(h)}$.
Both corrections clearly improve the variance, but the similarity of the degree  of improvement justifies our use of an effective device temperature.

\begin{figure}[h!]
\centering
\includegraphics[width=0.60\textwidth]{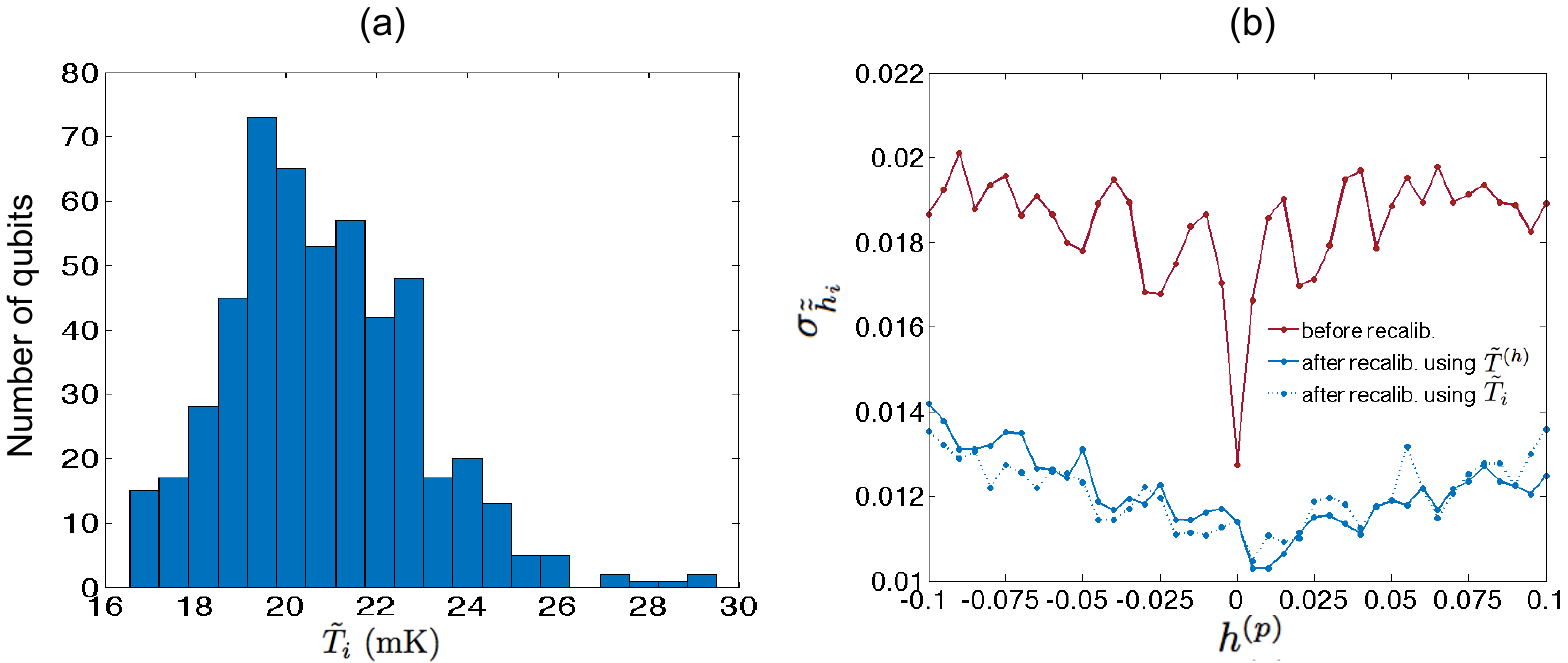}
\caption{
\textbf{Qubit temperatures and their effect.}
(a) The estimates of the individual qubit temperatures $\big\{\tilde{T}_i\big\}$ calculated from a single experiment without correction.
(b) The standard deviation $\sigma_{\tilde{\tilde{h}}_i}$ of the estimated field values $\tilde{\tilde{h}}_i$ for different values of $\hp$. 
}
\label{fig:TivsTtilde}
\end{figure}

\end{widetext}


\end{document}